\documentclass{PoS}

\usepackage{textgreek}
\usepackage{siunitx}
\DeclareSIUnit{\pe}{{p.e.}}
\usepackage{url}

\title{Commissioning and Performance of CHEC-S -- a compact high-energy camera for the Cherenkov Telescope Array}

\ShortTitle{Commissioning and Performance of CHEC-S}

\author{\speaker{J. J.~Watson}\\
        University of Oxford, Keble Road, Oxford OX1 3RH, UK\\
        Max-Planck-Institut f\"{u}r Kernphysik, P.O. Box 103980, D 69029 Heidelberg, Germany\\
        E-mail: \email{jason.watson@physics.ox.ac.uk}}
\author{J.~Zorn\\
        Max-Planck-Institut f\"{u}r Kernphysik, P.O. Box 103980, D 69029 Heidelberg, Germany\\
        E-mail: \email{justus.zorn@mpi-hd.mpg.de}}
\author{for the CTA GCT project\footnote{for collaboration list see PoS(ICRC2019)1177}\\
        URL: \email{https://www.cta-observatory.org/}\\}

\abstract{The Cherenkov Telescope Array (CTA) will present the next leap forward in gamma-ray astronomy, pushing beyond the present energy frontier to probe beyond 300 TeV. This capability is provided by the 70 Small Sized Telescopes (SSTs). The SSTs are spread across the four square kilometres of the array to detect the rare, but bright, Cherenkov showers produced by the highest-energy gamma rays. 

One proposed camera design for the SSTs is the Compact High Energy Camera (CHEC). Its compact and curved focal plane design is tailored for dual-mirror Schwarzschild-Couder telescopes, making it compatible with two of the three telescope proposals for the SSTs. The latest design of CHEC (known as CHEC-S) utilises silicon photomultipliers (SiPMs); an attractive alternative to traditional photomultiplier tubes, offering improved photon detection efficiency and photoelectron counting resolution for a large dynamic range, across tightly-packed pixels. However, SiPMs suffer from the phenomena of optical crosstalk, which degrades the ability to resolve the number of photons incident on the photosensor. CHEC-S also features full-waveform readout at nanosecond sampling resolution with a flexible trigger scheme. This is facilitated by the TARGET (TeV Array Read-out with GSa/s sampling and Event Trigger) modules attached to the SiPMs. 

This contribution describes the concept and technical design of CHEC-S and displays the key performance results, matched against the criteria required for a CTA camera. The limitation caused by the optical crosstalk of the SiPM is highlighted, and the expected performance with more recent iterations of the photosensor technology is also demonstrated.}

\FullConference{36th International Cosmic Ray Conference -ICRC2019-\\
		July 24th - August 1st, 2019\\
		Madison, WI, U.S.A.}

\begin{document}

\section{\label{01-introduction}Introduction} 

In order to push beyond the current energy frontier of gamma-ray astronomy, one must contend with the rapid drop off of gamma-ray flux with energy ($dN_\gamma/dE \sim E^{-2}$ or faster for typical sources). This can be achieved by increasing the effective gamma-ray detection area of an instrument. The Cherenkov Telescope Array (CTA) is the next generation of Imaging Atmospheric Cherenkov Telescope (IACT) array. It will provide a wide energy coverage while pushing beyond \SI{300}{TeV}. To achieve this goal, the southern site of the CTA will be realised with the following combination of three telescope tiers \cite{Acharya2013}:
\begin{itemize}
    \item 4 Large-Sized Telescopes (LSTs) -- with mirror diameter of \textasciitilde~\SI{23}{m} to
maximise the collection of Cherenkov photons from the low energy (but relatively frequent) showers (\SIrange{20}{150}{GeV}).
    \item 25 Medium-Sized Telescopes (MSTs) -- covering the mid-range energies between \SIrange{0.1}{10}{TeV}, with mirror diameters of \SI{12}{m}.
    \item 70 Small-Sized Telescopes (SSTs) -- spread over across the full 4 square kilometres of the array, increasing the number of detections of the rare high energy gamma-rays between \SIrange{1}{300}{TeV}. Only a small (\textasciitilde~\SI{4}{m}) mirror diameter is necessary to detect these high energy showers, enabling it to be cost effective to produce a large amount of this telescope tier.
\end{itemize}

To maximise the multiplicity of Cherenkov shower measurements, and enable the accurate reconstruction of showers at high impact distance, a large Field-of-View (FoV) is desirable. A dual-mirror Schwarzschild-Couder telescope design is capable of providing a \SI{9}{\degree} FoV while simultaneously reducing the plate scale by a factor of \textasciitilde~3 compared to single-mirror designs. This large reduction in plate scale allows for a compact and low-cost camera, for which novel opportunities in photosensor technology exist \cite{Vassiliev2007, White2017}. The Compact High Energy Camera (CHEC) has been designed to be compatible with all the dual-mirror SST proposals \cite{Sol2017, Maccarone2017}. The dual-mirror Cherenkov telescope design has been validated by the ASTRI project \cite{Giro2017}. These proceedings provide an overview of the commissioning activities and primary performance results of the second CHEC prototype, CHEC-S.

\section{\label{02-chec}The Compact High Energy Camera} 

\begin{figure}
	\centering
    \includegraphics[width=\textwidth]{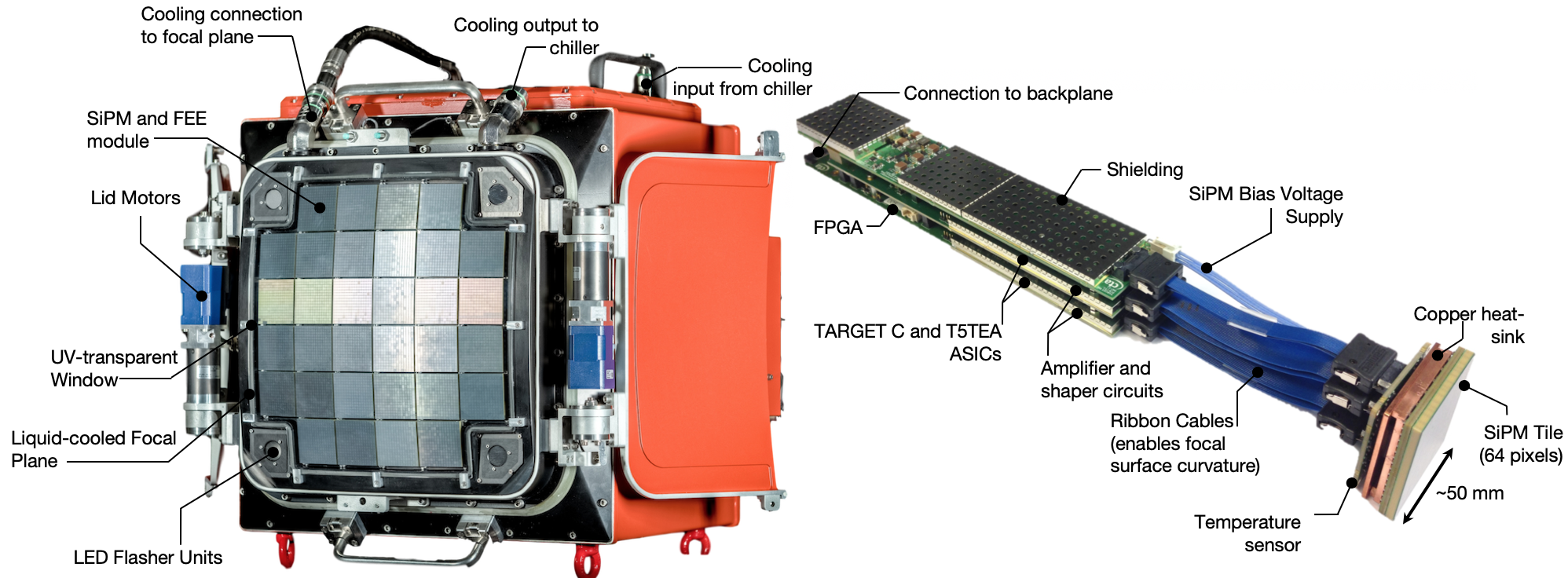} 
	\caption[Image of the CHEC-S focal surface.]{Left: Focal surface of the CHEC-S prototype, annotated with key components. Right: Image of the SiPM connected to the CHEC-S FEE with the components labelled. \cite{Watson2018}}
	\label{fig:camera}
\end{figure}

Two photosensor technologies have been explored for CHEC: 
\begin{itemize}
    \item CHEC-M, the first prototype, utilised Multi-Anode PhotoMultipliers (MAPMs) \cite{Zorn2017}.
    \item CHEC-S, the second prototype, utilises Silicon PhotoMultipliers (SiPMs).
\end{itemize}
Figure~\ref{fig:camera} displays the curved focal plane of the CHEC-S prototype. This curvature is required by the dual-mirror optics to prevent astigmatism \cite{Vassiliev2007}. CHEC has 2048 pixels in total spread over a \SI{0.4}{m} diameter. The SiPMs used in CHEC-S are the Hamamatsu S12642-1616PA-50 tiles. A single CHEC-S pixel is \textasciitilde$6\times6$~\si{mm^2}, comprised of four \textasciitilde$3\times3$~\si{mm^2} SiPM pixels. Each square grid of 64 pixels is connected to its own Front-End Electronics (FEE) module, performing readout from the photosensor via 64 channels. A backplane, not shown in Figure~\ref{fig:camera}, provides the power, clock, and trigger to the FEE modules \cite{White2017}. It also routes the data readout to the Data-Acquisition (XDACQ) board, which interfaces to the external camera server.

Aside from the components involved in the photosensor readout, CHEC has additional elements responsible for the upkeep of the camera. Thermal control of the camera is achieved with liquid cooling provided by an external chiller unit. In order to stabilise the temperature-sensitive SiPMs, the liquid is also circulated through the focal plane plate (via hollow ribs) \cite{White2017, Asano2018}. A curved window made from PMMA protects the SiPMs. LED flasher units in each of the four corners deliver a configurable illumination of the focal plane, via reflection off the telescope's secondary mirror, providing in situ calibration of the camera's photosensors. Finally, the camera lid protects the focal plane from the environment during downtime.

The remainder of this section will focus on the aspects of the camera that are relevant to, or directly influence, the performance results shown in Sections~\ref{03-intensityresolution} and \ref{04-triggerefficiency}.

\subsection{Photosensor}

An SiPM is a photon counting device. When a photon hits one of the >10,000 microcells in an SiPM pixel, an electron-hole pair is created which then produces an avalanche of excess charge carriers, resulting in a macroscopic current. The charge produced in this avalanche is limited by a quenching resistor, therefore each fired cell acts as a binary indication of an incident photon. This characteristic is responsible for the high photoelectron counting resolution of an SiPM, while the large number of cells allow for a large dynamic range. The other properties that make SiPMs an attractive photosensor for IACTs (as opposed to the traditional Photomultiplier Tubes) are \cite{Ghassemi2017}:
\begin{itemize}
    \item The higher Photon Detection Efficiency (PDE) attainable compared to traditional alkali-based photocathodes used for Photomultiplier Tubes.
    \item The lower voltage required for operation (on the order of \SIrange{10}{100}{V}).
    \item The mechanical reliability in terms of its ageing/warm-up.
    \item The higher fill factor of SiPM pixels, reducing dead space.
\end{itemize}

\begin{figure}[tb]
\centering
\begin{minipage}[t]{0.49\textwidth}
    \includegraphics[width=\textwidth]{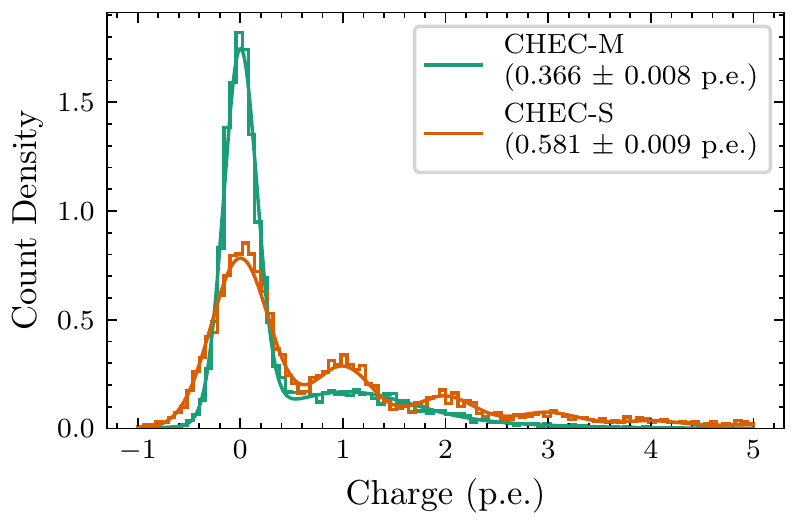}
    \caption{Comparison of single-photoelectron spectra between CHEC-M and CHEC-S for a single pixel, along with their corresponding fit function. Values in legend correspond to the average illumination in photoelectrons obtained from the fit. \cite{Watson2018}}
    \label{fig:spe}
\end{minipage}
\hfill
\begin{minipage}[t]{0.49\textwidth}
    \includegraphics[width=\textwidth]{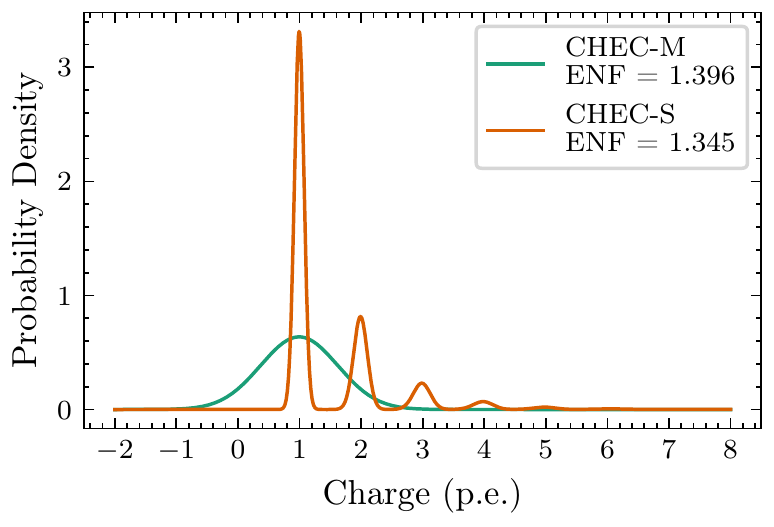}
    \caption{Probability density function for the measured value of charge when only a single photoelectron is initially generated in the photosensor (in the absence of electronic noise), compared between the MAPMs and SiPMs used in CHEC-M and CHEC-S respectively. \cite{Watson2018}}
    \label{fig:enf}
\end{minipage}
\end{figure}

While the photoelectron counting resolution of an SiPM is high, there is a phenomenon which unfortunately degrades the photon counting resolution known as Optical Crosstalk (OPCT). This occurs when secondary photons are generated inside the substrate material by the initial avalanche, which go on to fire adjacent cells. OPCT causes a larger average signal readout from the SiPM, but has the drawback of increasing the uncertainty in reconstructing the true number of initial fired cells (i.e. number of photons incident on the SiPM). Consequently, OPCT degrades the intensity resolution (described in Section~\ref{03-intensityresolution}). The OPCT can be lowered by reducing the overvoltage \cite{Asano2018}. However, reducing the overvoltage of an SiPM also reduces the gain and PDE, and therefore does not improve the intensity resolution. The SiPMs present in CHEC-S feature \textasciitilde \SI{33}{\percent} OPCT and \textasciitilde \SI{31}{\percent} PDE, as reported from the fit to the single-photoelectron spectrum (shown in Figure~\ref{fig:spe}), and the Hamamatsu data sheet\footnote{The PDE value was measured by Hamamatsu without the OPCT. The two values are disentangled.}.

Figure~\ref{fig:enf} describes the possible equivalent charges that can result from the multiplication of a singular photoelectron, using either an MAPM (used in CHEC-M) and an SiPM (from CHEC-S). These probabilities are inferred by the parameters extracted from the low-illumination measurements shown in Figure~\ref{fig:spe}. While the individual photoelectron (p.e.) peaks have a smaller width in the CHEC-S SiPMs, the OPCT in the SiPM produces additional higher photoelectron peaks. The Excess Noise Factor (ENF) is commonly used to quantify the spread in charge produced from the single photoelectron \cite{Watson2018, Teich1986}. The large OPCT present in these SiPM pixels results in a comparable ENF between the two photosensors used in the prototype cameras.

\subsection{Readout, Digitisation and Trigger} \label{sec:trigger}

The aforementioned FEE are all housed on what is known as a TARGET (TeV Array Readout with GSa/s sampling and Event Trigger) module, shown to the right in Figure~\ref{fig:camera}. The readout from the photosensor is performed by the four TARGET C ASICs, which sample 16 channels at a gigasample per second, storing the charge into a capacitor array with an equivalent maximum depth of \SI{16384}{ns} \cite{White2017, Funk2017}. Following a trigger signal from the backplane, the on-board Field Programmable Gate Array (FPGA) on every TARGET module instructs the TARGET C ASICs to look back inside its storage array to digitise the charges for all pixels into waveforms of a configurable length. Typically, the waveform length used in CHEC-S is 128 samples per pixel.

CHEC-S uses a two-level trigger logic \cite{Zorn2017, Zorn2019}:
\begin{enumerate}
    \item The signal of the 64 pixels on a module are input into the T5TEA ASIC onboard the TARGET module, in groups of $2\times2$ creating 16 ``superpixels''. The analogue sum of the signal in a superpixel is discriminated against a configurable threshold. If the sum surpasses the threshold, a first-level (L1) trigger is produced by the T5TEA ASIC.
    \item The L1 trigger signals are routed through the module to the backplane, resulting in 16 differential LVDS trigger signals per module and 512 ($32\times16$) in total for the whole camera. An FPGA on the backplane, referred to as the Trigger FPGA (TFPGA), accepts all 512 L1 trigger lines from the FEE modules. Implemented on the FPGA is a flexible camera-level trigger algorithm (programmable in the TFPGA firmware).
    \item Optimised for gamma-ray measurements with CHEC at TeV energies, the TFPGA is configured to a next-neighbour (NN) logic with a coincidence window of \SI{8}{ns}. If at least two L1 triggers, occurring within \SI{8}{ns}, are received at the TFPGA from neighbouring superpixels, a second-level (L2) trigger is formed. Given the camera geometry, there are 1910 possible NN superpixel combinations that may form an L2 trigger. Inter-module and diagonal neighbours are permitted.
    \item Following a successful L2 trigger, a readout request is sent to the FEE modules to initiate the full camera readout.
\end{enumerate}

\subsection{Commissioning}

CHEC-S was commissioned inside a dark room at the Max-Planck-Institut f\"{u}r Kernphysik, Heidelberg. A laser combined with a filter wheel containing a continuous neutral density filter and a diffuser were used to uniformly illuminate the camera. The average illumination level provided by different filter wheel settings were calibrated in terms of photoelectrons (and photons), delivering illuminations from the single photoelectron level, to beyond \SI{1000}{\pe}. This configurable illumination was used to calibrate the SiPM pixels, and investigate its performance in terms of intensity resolution and trigger efficiency.
\section{\label{03-intensityresolution}Intensity Resolution}

\begin{figure}
	\centering
    \includegraphics[width=\textwidth]{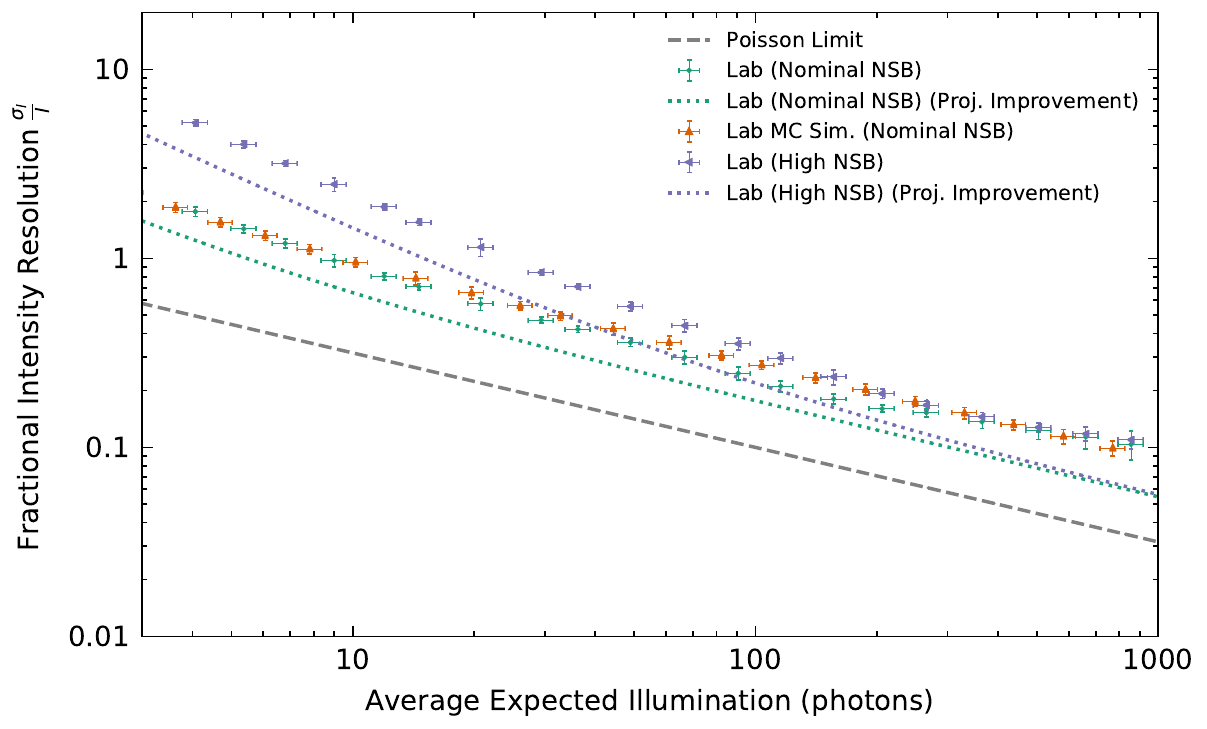} 
	\caption{Intensity resolution for CHEC-S lab measurements under nominal (\SI{40}{MHz}) and high (\SI{1}{GHz}) NSB conditions as compared to the lab MC simulation. The projections of the predicted improvements for the latest SiPM productions are also shown. The Poisson limit is displayed for reference.}
	\label{fig:res}
\end{figure}

The intensity resolution is used to evaluate how accurately the measured photon intensity $I_M$ represents the true photon intensity $I_T$ incident on the camera's pixels. Here we follow the standard CTA definition of intensity resolution:
\begin{equation} \label{eq:int_res}
\frac{\sigma_I}{I_T} = \frac{1}{I_T} \sqrt{\frac{\sum_{i=0}^N (I_{M_i} - I_T)^2}{N}},
\end{equation}
where $N$ is the number of measured charges, $I_{M_i}$, which are associated with that value of $I_T$ \cite{Watson2018}. Figure~\ref{fig:res} shows the intensity resolution results obtained by uniformly illuminating the camera in the lab, where a diffuse LED was used to control the NSB level.

Figure~\ref{fig:res} also displays the results obtained from a MC simulation which replicates this investigation, including a model of the CHEC-S prototype. The ``Lab MC Simulation'' points demonstrate that the intensity resolution performance of CHEC-S is reproducible in simulations, suggesting the MC model of the prototype is accurate. However, there are some small discrepancies in the comparison against MC that are still under investigation, likely due to aspects of the camera not included in the MC model, such as the electronic crosstalk and digitisation calibration \cite{Watson2018}. 

Lab investigations with reduced overvoltage, and simulations of sensors with lower OPCT, have both demonstrated that the largest source of uncertainty in intensity measurement for CHEC-S is the high OPCT of the SiPMs \cite{Watson2018}. These SiPMs are not intended to be used in the final camera design. The latest SiPM devices exhibit improved OPCT and PDE values, for example, \SI{8}{\percent} and \SI{39}{\percent} respective. Projections of the intensity resolution expected for CHEC-S with the improved SiPMs are also shown in Figure~\ref{fig:res}.
\section{\label{04-triggerefficiency}Trigger Efficiency} 

\begin{figure}[tb]
\centering
\begin{minipage}[t]{0.49\textwidth}
    \includegraphics[width=\textwidth]{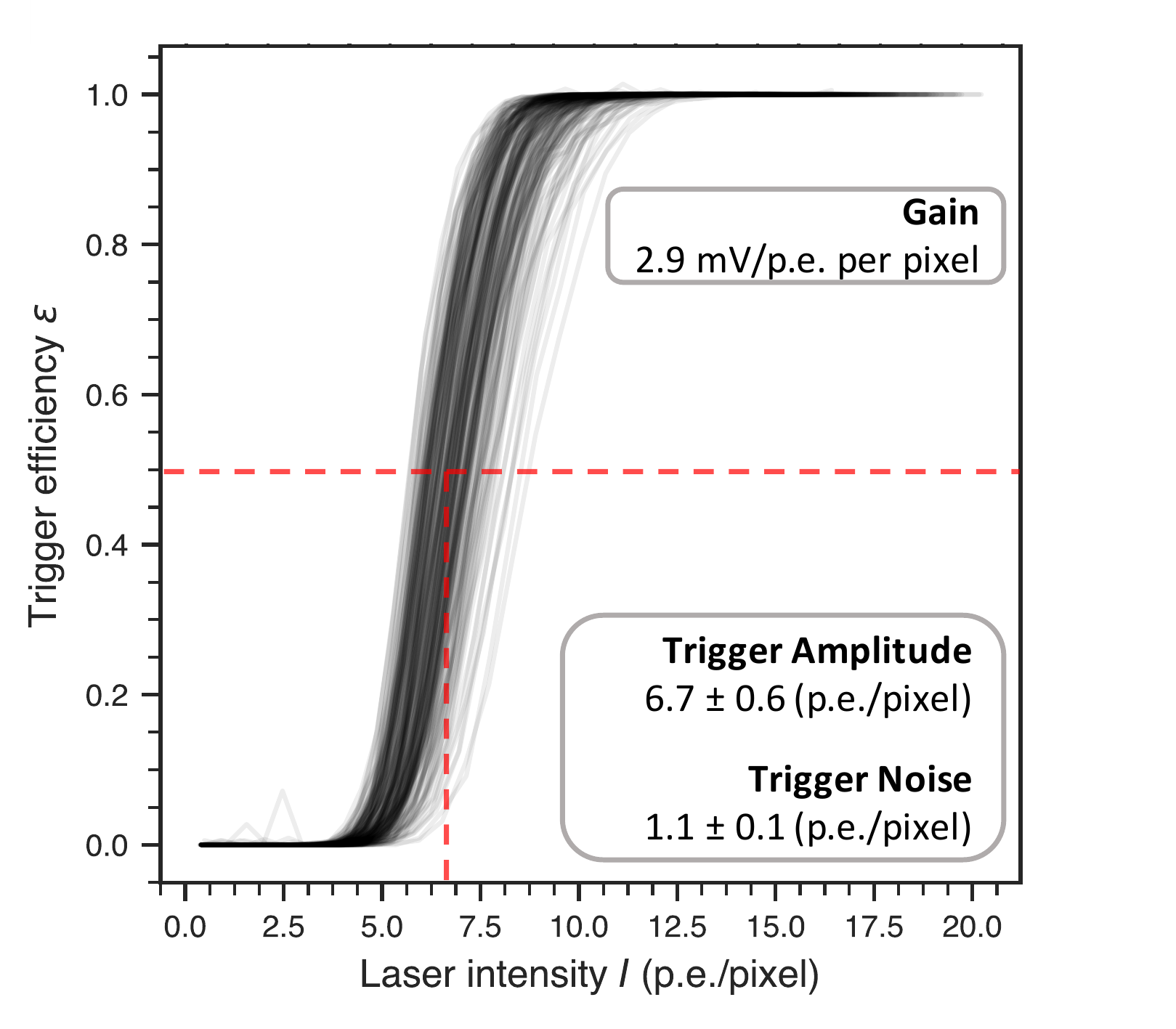}
    \caption{Efficiency of the L2 trigger as a function of illumination for 210 NN superpixel combinations. The \SI{50}{\percent} trigger efficiency is indicated by the horizontal red line, which defines the trigger amplitude (vertical red line).}
    \label{fig:trigger}
\end{minipage}
\hfill
\begin{minipage}[t]{0.49\textwidth}
    \includegraphics[width=\textwidth]{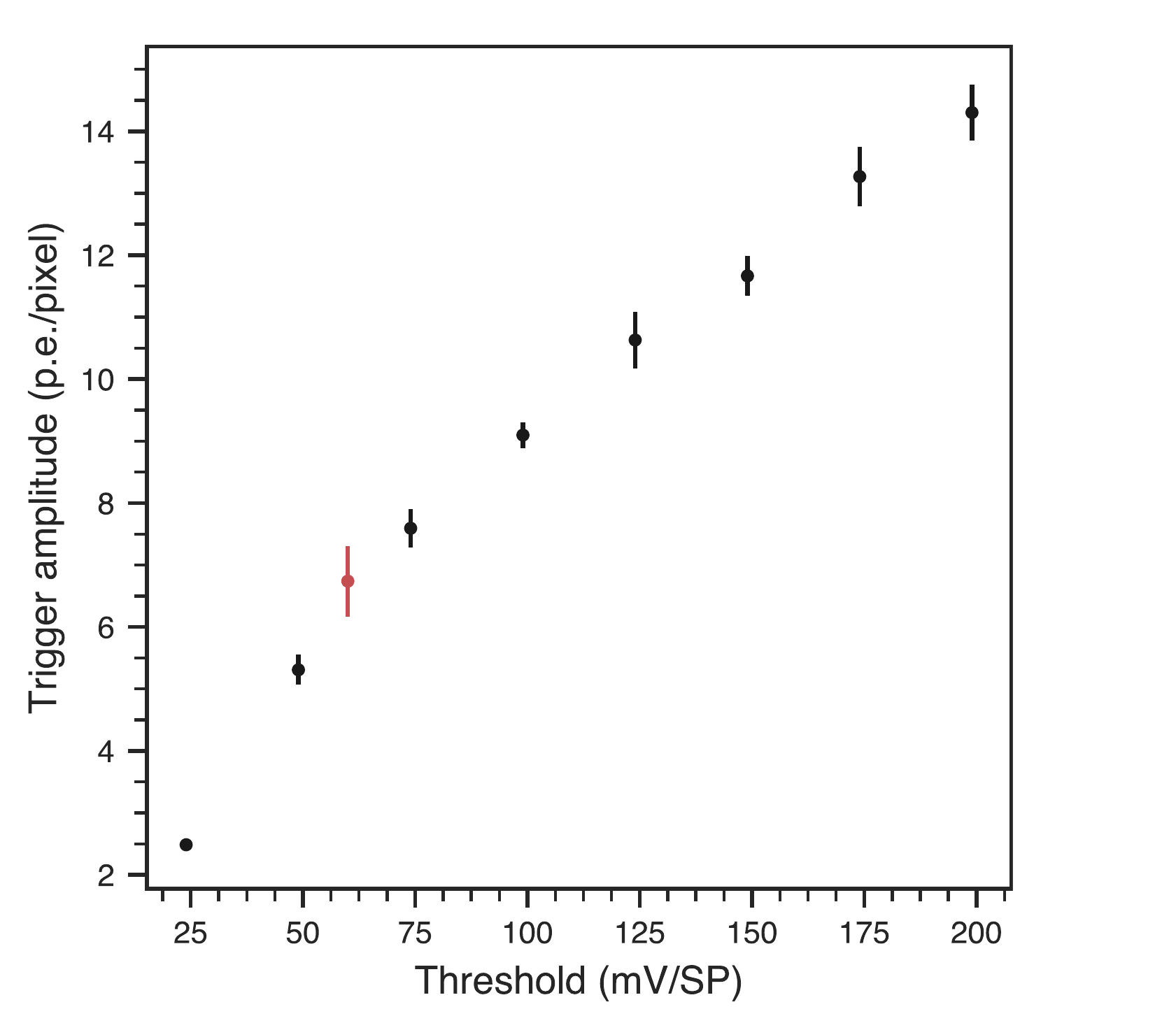}
    \caption{Trigger amplitude versus camera trigger threshold. The red point corresponds to the result shown in Figure~\ref{fig:trigger}.}
    \label{fig:trigger_thresh}
\end{minipage}
\end{figure}

By only enabling specific NN superpixel combinations, and iterating through illumination levels provided by the laser and filter wheel in the lab, one can examine the efficiency of the L2 trigger described in Section~\ref{sec:trigger} for each combination. Figure~\ref{fig:trigger} shows the trigger efficiency results for all 210 NN superpixel combinations contained in a subset of the camera (4 neighbouring modules). The investigation was conducted in the presence of \SI{60}{MHz} NSB and the camera trigger threshold set to \SI{60}{mV/superpixel}, with a gain per pixel of \SI{2.9}{mV/\pe}. The mean intensity at \SI{50}{\percent} efficiency, referred to the trigger amplitude, was measured to be \SI[separate-uncertainty = true]{6.7 \pm 0.6}{\pe/pixel}. This trigger amplitude is inline with that expected from the set threshold, and implies a uniformity across the full camera of \SIrange{10}{15}{\percent}. The uniformity can be improved further with an enhanced gain-matching calibration, as the voltage supplied to the SiPM is currently only configurable per superpixel. The trigger noise at \SI{50}{\percent} between combinations is also measured to be acceptable at \SI[separate-uncertainty = true]{1.1 \pm 0.1}{\pe/pixel}.

Figure~\ref{fig:trigger_thresh} displays the \SI{50}{\percent} trigger efficiency amplitude for other camera trigger thresholds, demonstrating its operational capability over a large range, which is desired for observations under varying NSB conditions. This relation can be utilised as a lookup to provide improved calibration of the trigger threshold, allowing for trigger determination in photoelectron units.
\section{\label{05-conclusion}Summary} 

Two key performance results of CHEC-S, the intensity resolution and trigger efficiency, have been demonstrated to match expectations. It is anticipated that the majority of components used in the CHEC-S prototype will also be used in the final production design of CHEC with the exception of the photosensors. More recent SiPM productions will provide lower OPCT and higher PDE, which will address the performance limits of the current prototype. CHEC-S has concluded its lab commissioning, and is now being tested on the ASTRI telescope prototype at Mt.~Etna, Sicily.

\acknowledgments{This work was conducted in the context of the CTA Consortium. We gratefully acknowledge support from the agencies and organisations listed under Funding Agencies at this website: \url{http://www.cta-observatory.org/consortium\_acknowledgments}}

\bibliographystyle{JHEP}
\bibliography{references}

\providecommand{\href}[2]{#2}\begingroup\raggedright\begin{thebibliography}{10}

\bibitem{Acharya2013}
B.~Acharya, M.~Actis, T.~Aghajani, G.~Agnetta, J.~Aguilar, F.~Aharonian et~al.,
  \emph{{Introducing the CTA concept}},
  \href{https://doi.org/10.1016/j.astropartphys.2013.01.007}{\emph{Astroparticle
  Physics} {\bfseries 43} (2013) 3}.

\bibitem{Vassiliev2007}
V.~Vassiliev, S.~Fegan and P.~Brousseau, \emph{{Wide field aplanatic two-mirror
  telescopes for ground-based $\gamma$-ray astronomy}},
  \href{https://doi.org/10.1016/j.astropartphys.2007.04.002}{\emph{Astroparticle
  Physics} {\bfseries 28} (2007) 10}.

\bibitem{White2017}
R.~White and H.~Schoorlemmer, \emph{{A Compact High Energy Camera (CHEC) for
  the Gamma-ray Cherenkov Telescope of the Cherenkov Telescope Array}},  in
  \emph{Proceedings of 35th International Cosmic Ray Conference --
  PoS(ICRC2017)}, (2017), p.~817,
  \href{https://doi.org/10.22323/1.301.0817}{DOI: 10.22323/1.301.0817}.

\bibitem{Sol2017}
H.~Sol, T.~Greenshaw, O.~{Le Blanc} and R.~White, \emph{{Observing the sky at
  extremely high energies with CTA: Status of the GCT project}},  in
  \emph{Proceedings of 35th International Cosmic Ray Conference --
  PoS(ICRC2017)}, (2017), p.~822,
  \href{https://doi.org/10.22323/1.301.0822}{DOI: 10.22323/1.301.0822}.

\bibitem{Maccarone2017}
M.~C. Maccarone, \emph{{ASTRI for the Cherenkov Telescope Array}},  in
  \emph{Proceedings of 35th International Cosmic Ray Conference --
  PoS(ICRC2017)}, (2017), p.~855,
  \href{https://doi.org/10.22323/1.301.0855}{DOI: 10.22323/1.301.0855}.

\bibitem{Giro2017}
E.~Giro, R.~Canestrari, G.~Sironi, E.~Antolini, P.~Conconi, C.~E. Fermino
  et~al., \emph{{First optical validation of a Schwarzschild Couder telescope:
  the ASTRI SST-2M Cherenkov telescope}},
  \href{https://doi.org/10.1051/0004-6361/201731602}{\emph{Astronomy {\&}
  Astrophysics} {\bfseries 608} (2017) A86}.

\bibitem{Watson2018}
J.~J. Watson, \emph{{Calibration and Analysis of the GCT Camera for the
  Cherenkov Telescope Array}}, Ph.D. thesis, University of Oxford, (2018),
  \href{https://ora.ox.ac.uk/objects/uuid:1794e10a-12b1-4f48-84df-582f48b0e702}{https://ora.ox.ac.uk/objects/uuid:1794e10a-12b1-4f48-84df-582f48b0e702}.

\bibitem{Zorn2017}
J.~Zorn, R.~White, J.~J. Watson, T.~P. Armstrong, A.~Balzer, M.~Barcelo et~al.,
  \emph{{Characterisation and testing of CHEC-M -- A camera prototype for the
  small-sized telescopes of the Cherenkov telescope array}},
  \href{https://doi.org/10.1016/j.nima.2018.06.078}{\emph{Nuclear Instruments
  and Methods in Physics Research, Section A: Accelerators, Spectrometers,
  Detectors and Associated Equipment} {\bfseries 904} (2018) 44}.

\bibitem{Asano2018}
A.~Asano, D.~Berge, G.~Bonanno, M.~Bryan, B.~Gebhardt, A.~Grillo et~al.,
  \emph{{Evaluation of silicon photomultipliers for dual-mirror Small-Sized
  Telescopes of Cherenkov Telescope Array}},
  \href{https://doi.org/10.1016/j.nima.2017.11.017}{\emph{Nuclear Instruments
  and Methods in Physics Research Section A: Accelerators, Spectrometers,
  Detectors and Associated Equipment} {\bfseries 912} (2018) 177}.

\bibitem{Ghassemi2017}
A.~Ghassemi, K.~Sato and K.~Kobayashi, \emph{{MPPC}},  technical note, accessed
  20/09/2018,
  \href{https://www.hamamatsu.com/resources/pdf/ssd/mppc{\_}kapd9005e.pdf}{https://www.hamamatsu.com/resources/pdf/ssd/mppc{\_}kapd9005e.pdf}.

\bibitem{Teich1986}
M.~Teich, K.~Matsuo and B.~Saleh, \emph{{Excess noise factors for conventional
  and superlattice avalanche photodiodes and photomultiplier tubes}},
  \href{https://doi.org/10.1109/JQE.1986.1073137}{\emph{IEEE Journal of Quantum
  Electronics} {\bfseries 22} (1986) 1184}.

\bibitem{Funk2017}
S.~Funk, D.~Jankowsky, H.~Katagiri, M.~Kraus, A.~Okumura, H.~Schoorlemmer
  et~al., \emph{{TARGET: A digitizing and trigger ASIC for the Cherenkov
  telescope array}},  in \emph{AIP Conference Proceedings}, vol.~1792, (2017),
  p.~080012, \href{https://doi.org/10.1063/1.4969033}{DOI: 10.1063/1.4969033}.

\bibitem{Zorn2019}
J.~Zorn, \emph{{CHEC -- A compact high energy camera for the Cherenkov
  Telescope Array}},
  \href{https://doi.org/10.1016/j.nima.2018.09.138}{\emph{Nuclear Instruments
  and Methods in Physics Research Section A: Accelerators, Spectrometers,
  Detectors and Associated Equipment} {\bfseries 936} (2019) 229}.

\end{thebibliography}\endgroup

\end{document}